# Investigating Effects of Perceived Technology-enhanced Environment on Self-regulated Learning: Beyond P-values

Chi-Jung Sui [1], Miao-Hsuan Yen [1 *], Chun-Yen Cha ng [1, 2, 3 *]

[1]*Graduate Institute of Science Education, National Taiwan Normal University, Taiwan*

[2]*Science Education Center, National Taiwan Normal University, Taiwan*

[3]*Department of Biology, Universitas Negeri Malang, Indonesia*

## Abstract

This study examined the effects of a technology-enhanced intervention on the self-regulation of 262 eighth-grade students, employing information and communication technology (ICT) and web-based self-assessment tools set against science learning. The data were analyzed using both maximum likelihood and Bayesian structural equation modeling to unravel the intricate relationships between self-regulation, self-efficacy, perceptions of ICT, and self-assessment tools. Our research findings underscored the direct and indirect impacts of self-efficacy, perceived ease of use, and perceived use of technology on self-regulation. The results revealed the predictive power of self-assessment tools in determining self-regulation outcomes, underlining the potential of technology-enhanced self-regulated learning environments. The study posited the necessity to transcend mere technology incorporation and to emphasize the inclusion of monitoring strategies explicitly designed to augment self-regulation. Interestingly, self-efficacy appeared to indirectly influence self-regulation outcomes through perceived the use of technology rather than direct influence. Analytically, this research indicated that Bayesian estimation could offer a more comprehensive insight into structural equation modeling by more accurately assessing our estimates' uncertainty. This research substantially contributes to comprehending the influence of technology-enhanced environments on students' self-regulated learning, stressing the importance of constructing practical tools explicitly designed to cultivate self-regulation.

Keywords: technology-enhanced environment, self-regulated learning, self-regulation, self-efficacy, Bayesian structural equation modeling

# 1. Introduction

As technology continues to advance, students increasingly utilize information and communication technology (ICT) and other technology-enhanced tools to aid their learning process. Hattie (2012) emphasized the importance of expert teachers identifying the most effective ways to represent their subject matter and fostering students' abilities to self-regulate, self-monitor, and self-evaluate their learning. Lim and Chan (2007) highlighted the evolving roles of teachers, shifting from knowledge receivers to knowledge constructors, and the transformation of technology's role from a mere learning tool to a facilitator of knowledge construction. Tondeur et al. (2016) conducted a comprehensive review of technology integration in education, discovering that schools not only tended to promote the use of ICT within a teacher-centered educational vision, particularly in mathematics and language classes but also promoted ICT as a student-centered educational approach, utilizing it as a presentation and communication tool. Undeniably, cultivating a technology-enhanced learning environment has emerged as a pivotal facet of 21st-century education, and understanding how to implement this optimally is absolutely critical.

Azevedo et al. (2004a) posited that technology-enhanced learning environments adopt the SRL framework as a scaffold to support students in science learning. In recent years, Taiwan government has collaborated with research institutions to develop guidelines for promoting SRL in junior high schools (Chen, 2021, 2022), emphasizing the significance of technology-enhanced self-regulated learning (SRL). Dignath et al. (2008) performed a meta-analysis of SRL interventions, concluding that SRL training positively impacted learning outcomes, strategy use, and motivation. The most significant intervention outcomes were achieved by combining different strategies. Moreover, the previous study found that employing a metacognitive strategy is a practical approach to learning (Cengiz-Istanbullu & Sakiz, 2022; Mathabathe &

Potgieter, 2014; Osterhage et al., 2019; Vemu et al., 2022). Self-assessment, a technique that enables students to evaluate their performance based on predetermined criteria, helps students identify their learning progress (Dang et al., 2018; Krebs et al., 2022; Zamora et al., 2016). As a result, a combination of self-assessment and a technology-enhanced environment might be a potential way to engage students in self-regulated learning.

Despite the growing body of literature on the effectiveness of ICT and self-assessment tools in fostering SRL (Lai et al., 2018; Panadero et al., 2017; Zheng, 2016), there remains a research gap in understanding the direct and indirect effects of various technology-enhanced interventions on students' post-intervention self-regulation. Therefore, our objective was to draw on the evidence gleaned from structural equation modeling (SEM) and construct an authentic, holistic, and interconnected representation of various factors that may not be uncovered using simpler statistical methods within the educational context. This study highlights the intricate relationship between technology-enhanced environments and students' self-regulation. In the present study, students were given an individual tablet to use Google to locate learning materials on teacher-assigned topics and access web-based self-assessment tools. We aimed to investigate how technology plays critical role in fostering students' SRL.

## 2. Literature review and research model

### *2.1 Self-regulated learning*

SRL has been a central focus in educational research for the past few decades due to its vital role in fostering academic success, which was defined as the self-directed process by which learners transformed their mental abilities into academic skills (Zimmerman, 2002). It could be traced back to earlier work on cognitive regulation, self-efficacy, and motivation (Bandura, 1982, 1986). The concept of SRL gained significant attention and recognition in the late 20th century (e.g., Pintrich & De Groot, 1990; Winne & Hadwin, 1998; Zimmerman, 1986), leading to the development of various SRL models. Panadero (2017) provided a comprehensive review of contemporary SRL frameworks. Some models, such as Zimmerman's socio-cognitive perspective (1986, 1990) and Winne and Hadwin's metacognitive perspective (1998), focused on cognition and metacognition. These models explored how learners set goals, monitored their progress, and adjusted their learning strategies based on feedback. Other models emphasized the role of motivation in SRL. Pintrich and his colleagues highlighted motivation, including goal orientation, task value, and self-efficacy, as critical factors in SRL (Pintrich & De Groot, 1990). Overall, these models contributed to a comprehensive understanding of the complex interplay between metacognition, cognition, and motivation in SRL.

*2.1.1 SRL in technology-enhanced environments*

As technology advances quickly and knowledge updates rapidly, SRL has become an essential component of technology-enhanced learning environments to foster students' learning of complex topics. Researchers have studied the role of adaptive scaffolding, SRL training, and technology-enhanced interventions in promoting SRL (Hong et al., 2021; Jansen et al., 2020; Lau & Jong, 2022; Raaijmakers et al., 2017; So et al., 2019; Yen et al., 2018; Zheng, 2016; Zhou et al., 2021). In science education, it could be traced back to around twenty years ago. Azevedo et al. (2004a) examined the effects of different scaffolding instructional interventions on students' ability to regulate their learning with hypermedia. They found that students in the adaptive scaffolding condition demonstrated better self-regulation by activating prior knowledge, monitoring their understanding, and engaging in adaptive help-seeking. Azevedo and his colleague conducted a follow-up study. Azevedo and Cromley (2004b) investigated the effectiveness of SRL training in facilitating college students' learning with hypermedia. They found that the SRL condition significantly facilitated the shift in learners' mental models. In recent years, Berglas-Shapiro et al. (2017) described developing a technology-enhanced SRL environment to foster students' SRL of complex science topics. The findings suggested that students could improve over time in regulating their learning and utilizing learning skills in a computerized system when provided with opportunities to practice, along with scaffolding. Additionally, Lai et al. (2018) conducted a quasi-experimental study on the effects of computer-supported self-regulation in science inquiry on learning outcomes, learning processes, and self-efficacy. The self-regulated science inquiry approach was found to improve student's learning achievement, especially for those students with higher self-regulation. Moreover, the students who conducted an inquiry with the SRL strategy increased their tendency of information help-seeking, self-efficacy, and several aspects of self-regulation, including time management, help-seeking, and self-evaluation. In conclusion, the previous studies highlighted the importance of technology-enhanced SRL among students. The researchers suggested incorporating these strategies can significantly improve students' learning outcomes, learning processes, and self-efficacy.

*2.1.2 SRL with self-assessment*

Within the SRL framework, self-assessment is a vital monitoring strategy that plays a crucial role in each phase (Winne & Hadwin, 1998; Zimmerman, 1986, 1990). For example, as students work on tasks, self-assessment stimulates progress monitoring and assists in making necessary adjustments to learning strategies. In the reflection phase, self-assessment encourages evaluating results based on predetermined criteria, allowing students to identify their strengths and disadvantages for improvement.

Science education researchers have explored the effects of providing students with self-assessment criteria, such as rubrics (Krebs et al., 2022; Safadi & Saadi, 2019), scripts (Zamora et al., 2016), and standards (Baars et al., 2014). Krebs et al. (2022) designed a rubric to assess abstract quality, aligned with four criteria introduced to students at the beginning of the lesson. The rubric's criteria included detailed instructions for specific aspects to consider across six levels. Krebs et al. (2022) found that rubrics improved accuracy when students wrote scientific abstracts and reduced cognitive load during the learning process. Safadi and Saadi (2019) used rubrics with varying scores for self-assessing geometric optics tasks. Their findings suggested that rubrics facilitated error detection strategies, improved self-assessment accuracy in solving physics problems, and enhanced academic achievement in secondary school physics courses. Similarly, scripts and standards were developed to support students in self-assessing their work using specific criteria. Consequently, educators and researchers often employ scaffolds, such as standards, checklists, or other evaluation tools, to enhance self-assessment judgments (Dunlosky et al., 2011; Lipko et al., 2009; Panadero & Jonsson, 2013). Ideally, learners should demonstrate accurate self-assessment, indicating strong metacognitive performance; therefore, we developed and implemented web-based self-assessment tools and standards (SAS) for students to self-evaluate their decision-making and argumentation in learning tasks.

*2.2 Perceived usefulness and ease of use*

Perceived usefulness (PU) and perceived ease of use (PEU) are the important factors of a well-established and influential theoretical framework for understanding and predicting user adoption of new technologies (Davis, 1989; Davis et al., 1989; Venkatesh & Davis, 1996, 2000). PU refers to the extent to which individuals believe that a specific technology can improve their work performance, while PEU reflects the degree to which individuals perceive the ease and effortlessness of using a particular technology. PEU directly affects PU, suggesting that users are more likely to find a valuable technology if they perceive it as easy to use. This relationship has been extensively explored in empirical research across diverse contexts and domains (Lee & Lehto, 2013; Sui et al., 2023; Teo, 2009). Over the years, researchers have examined the impact of PU and PEU on technology adoption in various fields, such as education, healthcare, and e-commerce. In educational settings, for instance, studies have shown that students' perceptions of the usefulness and ease of use of learning technologies are critical factors in determining their acceptance and usage (Lee & Lehto, 2013; Teo, 2009). A systematic review revealed two relatively underexplored aspects: the influence of external variables and the significance of various usage measures (Legris et al., 2003). Recent findings in social psychology, particularly concerning intrinsic motivation and self-efficacy, suggest that self-efficacy is a crucial determinant of an individual's behavior. Prior studies have investigated the impact of self-efficacy on perceived ease of use (Venkatesh & Davis, 2000; Yi & Hwang, 2003). In conclusion, PU and PEU were the critical factors of students' beliefs about the instruments. In this study, investigation of PU and PEU could provide valuable insights into understanding and predicting learners' adoption of technology-enhanced SRL.

### *2.3 Research model and hypotheses*

Theoretically, students' intrinsic motivation has been shown to impact their self-regulation (Pintrich & De Groot, 1990; Schunk, 1994). That is, students' perceptions of self-efficacy act as a determining factor in their motivation to employ SRL strategies. Bouffard-Bouchard et al. (1991) examined how self-efficacy affected the self-regulation of junior and senior high school students. They discovered that self-efficacy considerably impacted different aspects of self-regulation, regardless of cognitive ability and school grade differences. Based on these studies, we proposed two hypotheses:

H1: Self-efficacy will have a significant influence on the pre-test of initial self-regulation.

H2: Self-efficacy will have a significant influence on the post-intervention self-regulation.

Self-regulation is a key factor in SRL. Researchers have implemented learning activities in several studies to improve students' academic achievement and enhance their metacognition, such as self-regulation. Stanton et al. (2015) conducted a study in an extensive undergraduate biology course where students evaluated their learning strategies after the first and second exams. The results showed that almost all students were willing to reflect on and adjust their study plans accordingly. A subsequent qualitative study by Stanton et al. (2019) analyzed student responses to an exam self-evaluation task, revealing that both introductory and senior students demonstrated proficiency in evaluating their strategies. In addition, some students had an edge in assessing their overall plans and were more inclined to modify their strategies for improved effectiveness. Osterhage et al. (2019) conducted a study to investigate the impact of training on the accuracy of undergraduate self-assessment in biology courses. The training involved administering a pre-test on the first day of class and asking students to predict their performance on the next class day. Results showed that self-assessment training positively affected students' ability to accurately self-assess their performance. In other words, providing instruction on self-assessment can help improve metacognitive self-regulation.

H3: PreSR will have a significant influence on PostSR.

The use of technology was significantly influenced by two beliefs: perceived usefulness and perceived ease of use. Learners tended to view a system as more useful when it requires less effort, leading to perceived ease of use as a determinant of perceived usefulness. These relationships have been examined and supported by numerous previous studies (Davis, 1989; Davis et al., 1989; Lau & Jong, 2022; Lee & Lehto, 2013; Mayer & Girwidz, 2019; Teo, 2009; Venkatesh & Davis, 1996, 2000; Yang et al., 2019; Yi & Hwang, 2003). In this study, students acquired socio-scientific issues and argumentation skills using ICT and web-based self-assessment standards. Learners who strongly believe these instruments are easy to use may find them helpful in enhancing their learning effectiveness; thus, the following hypotheses were formulated.

H4: PEU of ICT (ICTPEU) will have a significant influence on PU of ICT (ICTPU).

H5: PEU of SAS (SASPEU) will have a significant influence on PU of SAS (SASPU).

The Social Cognitive Theory (Bandura, 1986) suggested that internal forces or external stimuli did not solely drive individuals but rather a dynamic interplay between the two. Self-efficacy, a crucial regulatory mechanism in this relationship, refers to people's judgments of their ability to complete specific tasks and influences their actions, effort, perseverance, and strategy selection in challenging situations (Venkatesh, 2000). Furthermore, self-efficacy represents an evaluation of users' ability to utilize technology, which may impact their perception of its usefulness and ease of use (Abdullah & Ward, 2016; Chahal & Rani, 2022; Gong et al., 2004). This study assessed students' self-efficacy within a technology-enhanced SRL context. Hypotheses were formulated to examine effects of self-efficacy on perceived usefulness and ease of use.

H6: Self-efficacy will have a significant influence on ICTPU.

H7: Self-efficacy will have a significant influence on SASPU.

H8: Self-efficacy will have a significant influence on ICTPEU.

H9: Self-efficacy will have a significant influence on SASPEU.

Numerous studies provide evidence that students' use of SRL strategies with technology was associated with belief in learning activities with technology. During the COVID-19 pandemic in China, Hong et al. (2021) conducted a study with a total of 531 participants. Their findings suggested that various aspects of self-regulated online learning—including task strategy, mood adjustment, self-evaluation, environmental structure, time management, and help-seeking—were associated with learners' perceived effectiveness of online learning. In a study by Cigdem (2015), the focus was on learners' self-regulation in blended learning environments. A total of 267 college students participated in a 15-week-long semester, learning computer programming through a blended approach, which combined face-to-face instruction with online learning. The findings revealed that perceived usefulness, satisfaction, and interactivity in the online learning environment significantly impacted learners' self-regulation. While PU and PEU have demonstrated their utility in diverse contexts, researchers recognized that external variables could also considerably impact those two factors. That is, external variables may directly or indirectly affect PU and PEU. Aligning with a recent meta-analysis, Legris et al. (2003) pinpointed a limitation of the model in not accounting for the role of external variables, prompting some studies to address this gap. Lau and Jong (2022) conducted an empirical study involving 1,381 secondary school students in Hong Kong. Their findings revealed that online SRL significantly predicted user acceptance of online learning. This study aimed to examine whether PU and PEU influenced students' self-regulation after a series of SSI and argumentation lessons based on technology-enhanced SRL.

H10: PreSR will have a significant influence on ICTPU.

H11: PreSR will have a significant influence on ICTPEU.

H12: PreSR will have a significant influence on SASPU.

H13: PreSR will have a significant influence on SASPEU.

H14: ICTPEU will have a significant influence on PostSR.

H15: ICTPU will have a significant influence on PostSR.

H16: SASPU will have a significant influence on PostSR

H17: SASPEU will have a significant influence on PostSR

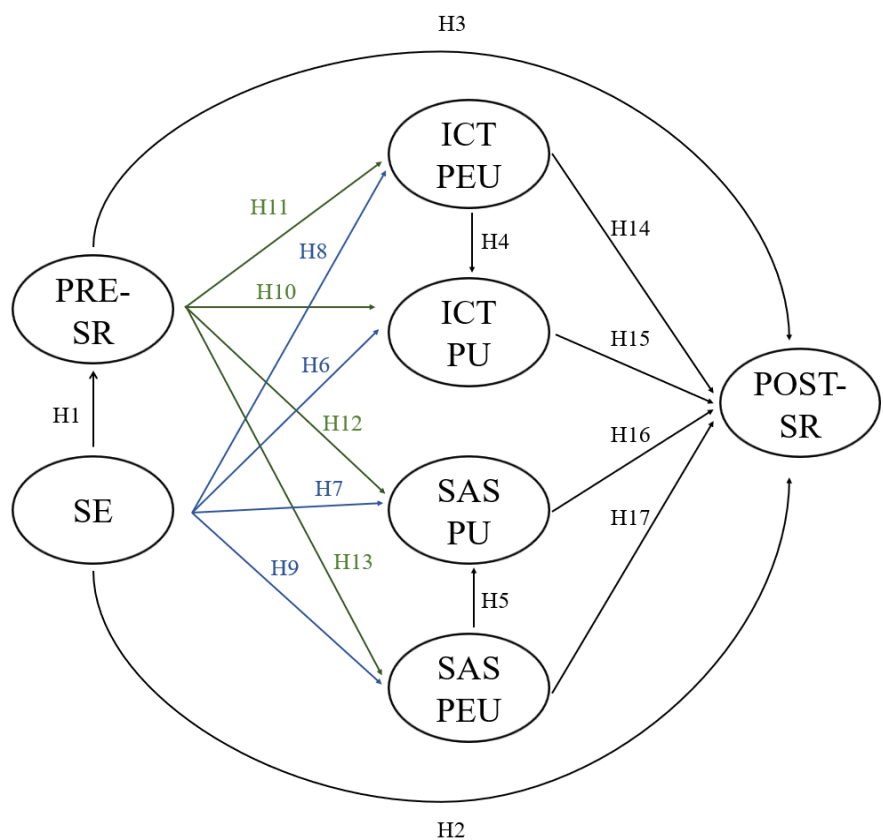

Figure 1

*The proposed model*

Overall, in this present study, we aimed to examine the proposed model (Figure 1), including the hypotheses (H1 - H17). By exploring these intricate relationships and the direct and indirect effects on post-intervention self-regulation, the study sought to gain a deeper understanding of which technology-enhanced interventions (i.e., ICT as well as self-assessment tools and standards) influenced students' self-regulation after lessons of SSI and argumentation.

## 3. Methods

### *3.1 Participants and settings*

The study involved 262 eighth-grade junior high school students from Northern Taiwan who had experienced several months of online learning due to the COVID-19 pandemic. During the distance learning, their teachers instructed them in SRL strategies. Following the end of remote learning, the students participated in eight lessons spanning two months, focusing on socio-scientific issues related to energy power stations. In the intervention, students were required to present claims, evidence, and reasoning (i.e., argumentation) for decision-making on energy issues. One example is the wind power issue:

> *"To expand renewable energy, the Taiwan Ministry of Economic Affairs has set a policy target of 20% renewable energy power generation by 2025. As a result, wind power generation is being actively promoted. Renewable energy is a form of green energy that does not produce carbon emissions during power generation. However, renewable energy often depends on natural resources. For instance, wind power generation requires a stable wind source for consistent power generation. There is extensive debate surrounding the pros and cons of wind power development, making it an energy choice worth discussing. Please state whether you agree or disagree with the decision to promote wind power generation and provide your argumentation."*

The students were given access to ICT instruments (e.g., tablets and Google). Students utilized the instruments to find the evidence and information to support them in decision-making and argumentation on power issues. Furthermore, they were required to use web-based self-assessment tools and standards for self-evaluating their tasks. For instance, while the student was writing their supporting arguments, they could refer to self-assessment tools alongside to self-monitor the quality of ideas and further regulate the behavior. Student's operations and responses were logged in the system, and students further received the logs and individual feedback from their instructors. This environment placed participants in a technology-enhanced SRL context.

*3.2 Survey instrument*

The metacognitive self-regulation construct was adapted from the Motivated Strategies for Learning Questionnaire (MSLQ) developed by Pintrich (1991). The adapted construct consists of five items tailored to the context of learning socio-scientific issues and argumentation. It was administered pre- and post-test to assess students' perceived metacognitive self-regulation. The Cronbach's alpha values for PreSR and PostSR were 0.87 and 0.90, indicating good internal consistency for both measures. The self-efficacy construct was adapted from the Motivated Strategies for Learning Questionnaire (MSLQ) developed by Pintrich (1991). The adapted construct consists of three items designed to measure students' self-efficacy in technology-enhanced SRL, including using tablets, Google, and standards to self-evaluate the quality of their argumentation. The Cronbach's alpha was 0.92, reflecting strong internal consistency. The perceived usefulness and ease of use scales were adapted from Davis (1989), resulting in the development of the constructs. Two constructs were designed to measure the perceived usefulness and ease of use of SAS, which helps students self-assess their argumentation quality. On the other hand, another two constructs were used to measure ICT's perceived usefulness and ease of use. Each construct consisted of three items.. The Cronbach's alpha values for SASPU and SASPEU were 0.89 and 0.90 and for ICTPU and ICTPEU were 0.86 and 0.92, indicating good internal consistency for all measures. All items were rated on a 7-point Likert scale ranging from 1 (strongly disagree) to 7 (strongly agree), shown in Table1A.

*3.3 Analytic strategies*

Bayesian statistics is an approach to data analysis grounded in Bayes' theorem. In this approach, the background knowledge is expressed as a prior distribution, which is then combined with the observed data through a likelihood function; therefore, Bayesian statistics involves updating the available knowledge about parameters in a statistical model using observed data. By integrating these components, the posterior distribution is determined. The posterior distribution provides valuable insights and can be used to predict future events (Kruschke & Liddell, 2018; McElreath, 2020; van de Schoot et al., 2021). Bayesian SEM (B-SEM) offers distinct advantages: it (1) does not rely on asymptotic theory, which assumes a normal distribution of parameter estimates based on large-sample theory, and (2) enables a transparent interpretation of the posterior distribution, providing valuable insights into the most credible values of parameters and their associated uncertainty. This interpretability fosters a comprehensive understanding of the underlying model and its implications. When applied to small samples, b-SEM (3) demonstrates superior performance (Muthén & Asparouhov, 2012; Smid et al., 2020; van de Schoot et al., 2017).

The hypothesized relationships among the variables were examined using maximum likelihood estimation with the lavaan package (version 0.6-15) in R (Rosseel, 2012). The data analysis process comprised two main steps. First, the maximum likelihood (ML) and Bayesian estimations were conducted to estimate the measurement model, which assessed the ability of the observed variables to capture the underlying constructs. Second, we used ML-SEM to test the structural relationships, which outlined the proposed relationships between the factors. Additionally, the B-SEM was used to capture the uncertainty of ML-SEM with the blavaan package (version 0.4-7) in R (Merkle & Rosseel, 2018) because B-SEM could provide a full probability distribution for each model parameter, which allowed for a more comprehensive assessment of uncertainty. The Bayesian model was estimated with three independent MCMC chains with Hamiltonian Monte Carlo sampling (Duane et al., 1987). The estimation was performed with 20,000 iterations for each chain after a burn-in period of 20,000 iterations. Non-informative priors for estimation were specified for the factor loadings with normal(0, 2), variances with normal(1,0.5), and regressions with normal(0,2). When the sample size is large and vague priors are specified, the posterior distribution is predominantly informed by the likelihood function, and results become asymptotically equal to a ML solution (Garnier-Villarreal & Jorgensen, 2020).

## 4. Results

### *4.1 Descriptive statistics and difference of SR, PU, and PEU*

Table 1 shows each factor's mean, standard deviation, skewness, kurtosis, and Cronbach's alpha. According to Kline's (2011) suggestions, the skewness and kurtosis indices should be within |3| and |10|, respectively. The skewness values for the variables ranged from -0.57 to 0.13, while the kurtosis values ranged from -0.65 to -0.03. The results in the present study met the assumption of a normal distribution, warrant further investigation of the measurement model.

Table 1

*The results of descriptive statistics*

|        | Mean | Std. Deviation | Skewness | Kurtosis | alpha |
|--------|------|----------------|----------|----------|-------|
| PreSR  | 4.11 | 1.29 | 0.06  | -0.03 | 0.87 |
| PostSR | 4.41 | 1.31 | -0.10 | -0.13 | 0.90 |
| SE     | 4.09 | 1.48 | 0.13  | -0.65 | 0.92 |
| ICTPU  | 5.13 | 1.42 | -0.37 | -0.34 | 0.86 |
| ICTPEU | 5.39 | 1.33 | -0.57 | -0.17 | 0.92 |
| SASPU  | 4.55 | 1.39 | -0.26 | -0.27 | 0.89 |
| SASPEU | 4.65 | 1.34 | -0.24 | -0.11 | 0.90 |

Note: SE stands for self-efficacy.

## *4.2 Analysis of the measurement model*

The entire measurement model was statistically tested using the lavaan (version 0.6-15) and blavaan (version 0.4-7) packages in R. The analysis used both the maximum likelihood and Bayesian estimation with a sample size of 262 observations to examine the measurement model of the latent variables: PreSR, PostSR, self-efficacy, ICTPU, ICTPEU, SASPU, and SASPEU. Composite reliability (CR) and average variance extraction (AVE) were used to establish the reliability and validity of observed variables in the model, and the minimum acceptable AVE and CR values are .500 and .600, respectively, or above (Fornell & Larcker, 1981; Kline, 2011). As shown in Table 2A, all CR values are above the recommended threshold of 0.7, indicating good internal consistency for each construct. Additionally, all AVE values are above the threshold of 0.5, which suggests adequate convergent validity for each construct.

The results of the maximum likelihood estimation supported the adequacy of the model fit ($\chi2$ = 462.27, $\chi2/df$ = 1.82, $p < .001$, TLI = .960, CFI = .953, RMSEA = .056, SRMR = .047). Parameter estimates were statistically significant, including factor loadings and covariance estimates ($p < 0.001$). Based on the Bayesian estimation, the goodness of fit was assessed using Bayesian versions of the following fit indices:

RMSEA, CFI, TLI, and GammaHat. The results of the Bayesian estimation supported the adequacy of the model fit (PPP < .001, TLI = .946 [0.941, 0.951], CFI = .960 [0.956, 0.964], RMSEA = .059 [0.057, 0.062], GammaHat = .940 [0.934, 0.945]). In contrast, the posterior predictive *p*-value (PPP) was less than .001, indicating a poor fit. The PPP, which ranges from 0 to 1, has been commonly used as a fit statistic in Bayesian modeling with a value of 0.5 indicating a good fit. However, recent research has suggested that the PPP may not be a reliable fit indicator, particularly in complex models (Asparouhov & Muthén, 2021; Garnier-Villarreal & Jorgensen, 2020). In conclusion, both estimations' fit indices revealed the measurement model's good fit (Asparouhov & Muthén, 2021; Garnier-Villarreal & Jorgensen, 2020; Hu & Bentler, 1999).

*4.3 Analysis of the structural model*

Significant positive correlations between the factors were identified (Table 2), with Pearson correlation coefficients ranging from 0.301 to 0.718. The Bayes factor ($BF_{10}$) values are used to compare the strength of evidence for whether there is a correlation or not (Ly et al., 2018; Ly et al., 2016); namely, $BF_{10}$ represents the ratio of the likelihood of that there is a correlation to that there is no correlation. The larger the $BF_{10}$, the more substantial the evidence for a correlation between each factor. The results showed that the values of $Log(BF_{10})$ ranged from 9.71 to 125.63, supporting the strong presence of a correlation over no correlation at all. These findings suggested that the relationships among the factors warrant further SEM investigation.

Table 2

*Pearson's correlation of each factor*

| Factor | | PreSR | PostSR | SE | ICTPU | ICTPEU | SASPU |
|---|---|---|---|---|---|---|---|
| PostSR | r | 0.55 | | | | | |
| | $Log(BF_{10})$ | 43.97 | | | | | |
| SE | r | 0.66 | 0.52 | | | | |
| | Log(BF$_{10}$) | 72.41 | 38.41 | | | | |
| ICTPU | r | 0.39 | 0.39 | 0.46 | | | |
| | Log(BF$_{10}$) | 18.84 | 18.66 | 28.09 | | | |
| ICTPEU | r | 0.30 | 0.30 | 0.44 | 0.71 | | |
| | Log(BF$_{10}$) | 9.71 | 9.79 | 25.65 | 89.56 | | |
| SASPU | r | 0.47 | 0.80 | 0.49 | 0.42 | 0.32 | |
| | Log(BF$_{10}$) | 29.03 | 130.15 | 32.11 | 22.54 | 11.52 | |
| SASPEU | r | 0.38 | 0.72 | 0.45 | 0.41 | 0.33 | 0.79 |
| | Log(BF$_{10}$) | 17.75 | 91.06 | 27.15 | 21.34 | 12.43 | 125.63 |

The structural model of the hypothesized relationships was statistically tested using the lavaan (version 0.6-15) and blavaan (version 0.4-7) packages in R. The results of ML-SEM indicated a good fit of the structural model ($\chi2$ = 478.42, $\chi2$/df = 1.84, *p* < .001, TLI = .952, CFI = .959, RMSEA = .057, SRMR = .063). On the other hand, the results of the B-SEM analysis were obtained with three MCMC chains of 40,000 iterations each, including 20,000 burn-in iterations. Uninformative priors were used for all parameters. The goodness of fit was assessed using Bayesian versions of the following fit indices: RMSEA, CFI, TLI, and GammaHat. The results of ML-SEM indicated a good fit of the structural model (PPP < .001, TLI = .944 [0.939, 0.949], CFI = .958 [0.954, 0.962], RMSEA = .061 [0.058, 0.063], GammaHat = .937 [0.931, 0.942]). In contrast, the PPP was less than .001, indicating a poor fit; however, this index may not be reliable in the complex model. All Rhat values were at the optimal value of 1.000, indicating good convergence of the model. The adequate effective sample size ranged from 37,119 to 55,355, signifying sufficient precision in the parameter estimates. Based on these indices, the model adequately fitted the data. In summary, the results suggested that the proposed model demonstrates a good fit to the observed data, as evidenced by the acceptable values of various fit indices (Asparouhov & Muthén, 2021; Garnier-Villarreal & Jorgensen, 2020; Hu & Bentler, 1999).

Based on both ML-SEM and Bayesian SEM results (Table 3), self-efficacy strongly influenced PreSR (H1). Self-efficacy did not strongly influence PostSR (H2). PreSR strongly influenced PostSR (H3). ICTPEU strongly influenced ICTPU (H4). SASPEU strongly influenced SASPU (H5). Self-efficacy strongly influenced ICTPEU and SASPEU (H8 and H9). PreSR did not strongly influence ICTPEU and SASPEU (H10 and H13). PreSR, however, strongly influenced both ICTPU and SASPU (H11 and H12). ICTPEU and ICTPU did not strongly influence PostSR (H14 and H15). SASPU strongly influenced PostSR (H16). SASPEU had no significant influence on PostSR ($p= .102$); however, based on the Bayesian SEM results ($\beta = .18$, 95% CI [-.03, .40]), SASPEU had a marginally strong influence on PostSR (H17). Still, this relationship was not as strong as the influence of SASPU. Note: CI stands for confidential interval, pd stands for probability of direction, and ESS stands for effective sample size.

Table 4

Indirect effects on PostSR

| Path | Estimate |
| --- | --- |
| SE → PreSR → PostSR | 0.15 |
| SE → PreSR → SASPU → PostSR | 0.07 |
| SE → SASPEU → PostSR | 0.08 |
| SE → SASPEU → SASPU → PostSR | 0.22 |

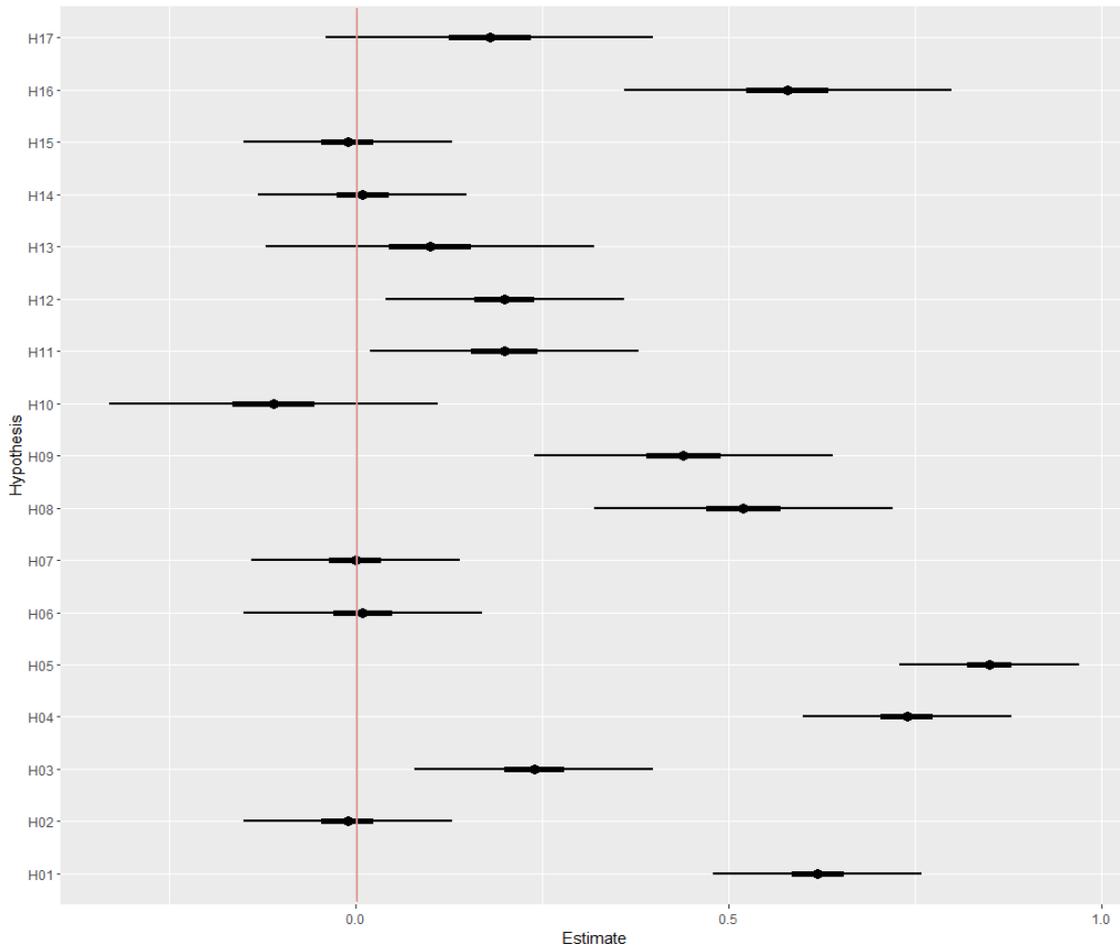

Figure 2 shows the distribution of each hypothesis estimate. PostSR was found to be determined by PreSR, SASPU, and SASPEU, resulting in $R^2$ of 0.82 (both ML-SEM and B-SEM). The results indicated that the variance in PostSR could be powerfully explained by these predictor factors, suggesting that these factors had a strong relationship with PostSR and played an essential role in determining students' PostSR.

Our examination of the regression results and predefined hypotheses focused on elucidating the direct and indirect effects on PostSR. The results demonstrated direct impacts of PreSR, SASPU and SASPEU on PostSR. The indirect effects on PostSR were discernible through additional variables within the model. We identified four indirect paths: self-efficacy → PreSR → PostSR, self-efficacy → PreSR → SASPU→ PostSR, self-efficacy → SASPEU → PostSR, and self-efficacy → SASPEU → SASPU

→ PostSR (Table 4). Even though the relationship between SASPEU and PostSR was only supported by the B-SEM results, we investigated this potential path further. The affordance of B-SEM piqued our interest in providing extensive evidence of structural relations. Based on product-of-coefficients strategy (Preacher & Hayes, 2008), the total indirect effect of self-efficacy on PostSR was 0.52. Our findings indicated that these indirect paths positively contribute to the overall indirect effect of self-efficacy on PostSR. In conclusion, our analysis unveiled the significance of both direct and indirect impacts on PostSR, underlining the critical roles of self-efficacy, PreSR, SASPU, and SASPEU in shaping PostSR outcomes.

Table 3

The results of ML-SEM and B-SEM

| Hypothesis | ML-SEM | | | B-SEM | | | | | | |
|---|---|---|---|---|---|---|---|---|---|---|
| | Coefficient | p | Supported By ML-SEM | Median | Posterior SD | 95% CI | | pd | Rhat | ESS |
| H1: SE → PreSR | .72 | <.001 | Yes | .62 | .07 | [ .49 | .75] | 100% | 1.000 | 37119 |
| H2: SE → PostSR | -.01 | .890 | No | -.01 | .07 | [-.15 | .12] | 57% | 1.000 | 48785 |
| H3: Pre-SR → PostSR | .22 | .006 | Yes | .24 | .08 | [ .09 | .39] | 100% | 1.000 | 47456 |
| H4: ICTPEU → ICTPU | .71 | <.001 | Yes | .74 | .07 | [ .61 | .89] | 100% | 1.000 | 52492 |
| H5: SASPEU → SASPU | .81 | <.001 | Yes | .85 | .06 | [ .73 | .98] | 100% | 1.000 | 53409 |
| H6: SE → ICTPU | .01 | .894 | No | .01 | .08 | [-.15 | .17] | 56% | 1.000 | 52918 |
| H7: SE → SASPU | .00 | .985 | No | .00 | .07 | [-.14 | .13] | 52% | 1.000 | 55173 |
| H8: SE → ICTPEU | .58 | <.001 | Yes | .52 | .10 | [ .34 | .72] | 100% | 1.000 | 49066 |
| H9: SE → SASPEU | .45 | <.001 | Yes | .44 | .10 | [ .25 | .64] | 100% | 1.000 | 50352 |
| H10: PreSR → ICTPEU | -.10 | .308 | No | -.11 | .11 | [-.33 | .10] | 84% | 1.000 | 54336 |
| H11: PreSR → ICTPU | .19 | .028 | Yes | .20 | .09 | [ .04 | .38] | 99% | 1.000 | 55055 |
| H12: PreSR → SASPU | .17 | .007 | Yes | .20 | .08 | [ .05 | .36] | 100% | 1.000 | 55355 |
| H13: PreSR → SASPEU | .09 | .388 | No | .10 | .11 | [-.12 | .33] | 82% | 1.000 | 53510 |
| H14: ICTPEU → PostSR | .01 | .947 | No | .01 | .07 | [-.14 | .15] | 53% | 1.000 | 50849 |
| H15: ICTPU → PostSR | -.01 | .872 | No | -.01 | .07 | [-.15 | .13] | 57% | 1.000 | 53993 |
| H16: SASPU → PostSR | .62 | <.001 | Yes | .58 | .11 | [ .37 | .80] | 100% | 1.000 | 47493 |
| H17: SASPEU → PostSR | .19 | .102 | No | .18 | .11 | [-.03 | .40] | 96% | 1.000 | 49161 |

Note: CI stands for confidential interval, pd stands for probability of direction, and ESS stands for effective sample size.

Table 4

Indirect effects on PostSR

| Path | Estimate |
| --- | --- |
| SE → PreSR → PostSR | 0.15 |
| SE → PreSR → SASPU → PostSR | 0.07 |
| SE → SASPEU → PostSR | 0.08 |
| SE → SASPEU → SASPU → PostSR | 0.22 |

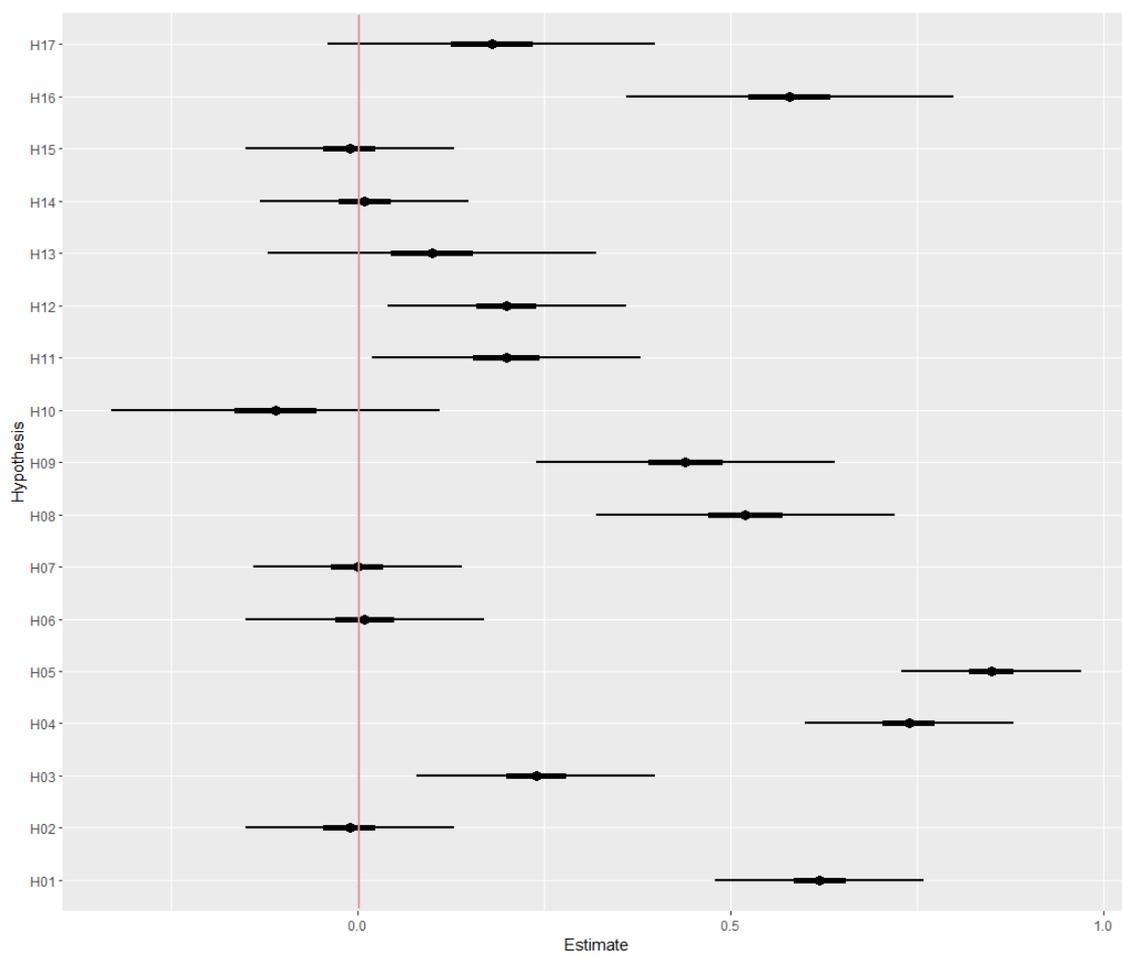

Figure 2

*The posterior estimate of each hypothesis*

Note: Dots are medians, thin lines are 95% CI, and thick lines are 50% CI.

## 5. Discussion

We aimed to examine what factors could determine students' self-regulation outcomes through both maximum likelihood and Bayesian estimations. As shown in Figure 3, the results of this study showed that initial self-regulation, perceived usefulness, and perceived ease of use of web-based self-assessment standards had direct effects on self-regulation outcomes. Moreover, self-efficacy affected self-regulation outcomes indirectly. Initial self-regulation, perceived usefulness and perceived ease of use of self-assessment standards served as the intermediary factors, bridging the associations between these indirect effects. Overall, we found evidence that the resulting measurement and structural models revealed a good fit to observe the relationships and supported existing theories and assumptions that such factors affected students' self-regulation outcomes.

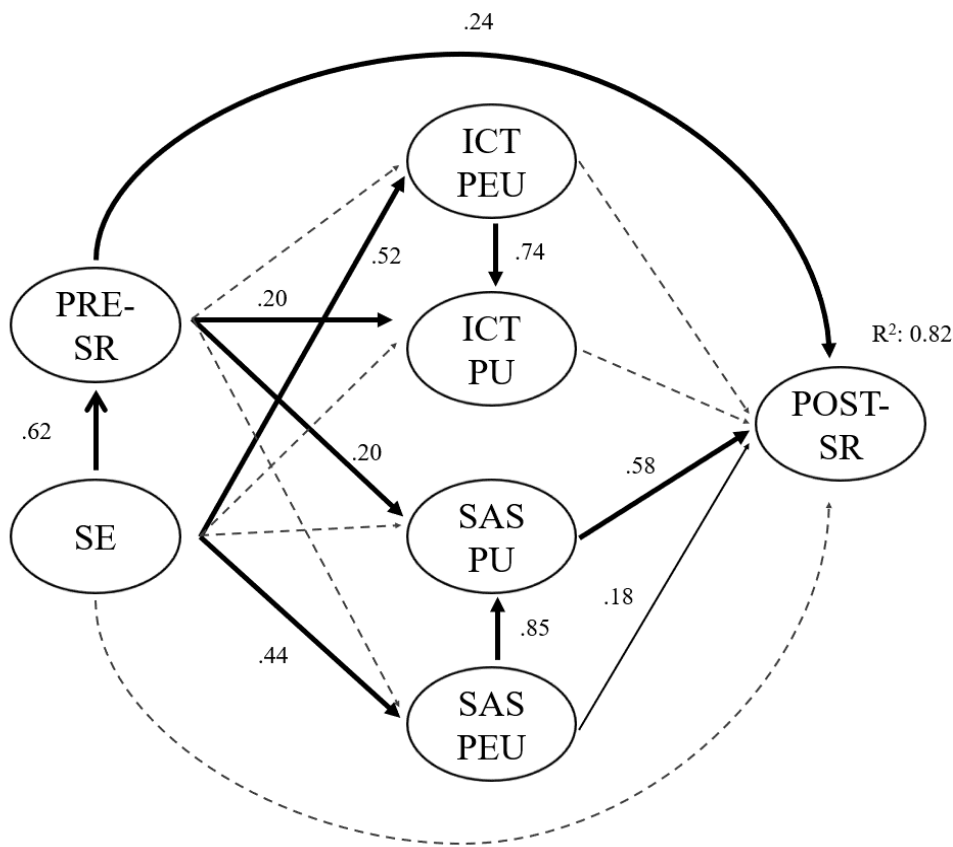

Figure 3

*The structural model based on B-SEM.*

Note: The thick arrows stand for strong influence, the thin arrows stand for moderate influence, and the dashed arrows stand for weak or no influence.

Our results revealed a correlation between self-efficacy and PreSR and PostSR. However, while self-efficacy significantly influenced initial self-regulation (H1), it did not appear to impact self-regulation following the intervention (H2), as indicated by our model. The results suggested that the role of self-efficacy was integral in shaping initial self-regulation, implying that learners with higher levels of confidence in technology-enhanced environments tended to own better self-regulation at the outset. This relationship marked the significance of self-efficacy in cultivating an learner's ability to manage their learning process effectively in line with the previous research (Bouffard-Bouchard et al., 1991; Pintrich & De Groot, 1990; Schunk, 1994). However, our study indicated that self-efficacy did not influence self-regulation after the intervention. This compelling finding suggested that while self-efficacy was crucial in establishing initial self-regulation, other factors became more influential in maintaining or improving self-regulation once the intervention had been implemented. In this study, we speculated that it might be that once individuals have engaged with the intervention and enhanced their initial self-regulation abilities, their self-efficacy played a less dominant role as they relied more on the technology-enhanced environment factors. For instance, SAS's perceived usefulness and ease of use might become more impactful in this stage. Although there was a lack of a strong direct relationship between self-efficacy and PostSR, we found the indirect effects of self-efficacy on PostSR: SE → PreSR → PostSR, SE → PreSR → SASPU→ PostSR, SE → SASPEU → PostSR, and SE → SASPEU → SASPU → PostSR. These findings emphasized the importance of considering the multifaceted self-regulation variables in technology-enhanced learning environments. We should fully understand the dynamics at play and identify other potential factors that might impact self-regulation after an intervention.

As we expected consistency with the previous studies (Osterhage et al., 2019; Stanton et al., 2019; Stanton et al., 2015), the early self-regulation capacities served as a reliable indicator of future self-regulation outcomes (H3), suggesting that learners who already possessed strong self-regulation abilities before an intervention was likely to maintain or even enhance these abilities afterward. Additionally, students perceived usefulness of both technology-enhanced instruments could be determined by perceived ease of use (H4 and H5), suggesting that if students found a tool easy to use, they were more likely to see it as beneficial and incorporate it into their learning process. These findings were in line with the previous research (Lau & Jong, 2022; Lee & Lehto, 2013; Mayer & Girwidz, 2019; Teo, 2009; Venkatesh & Davis, 2000; Yang et al., 2019; Yi & Hwang, 2003) and highlighted that ease of use directly impacted the user's perceptions of tools' effectiveness in enhancing learning.

We found that self-efficacy (i.e., students believed they could learn well in a technology-enhanced environment) acted as a critical determinant of ease of use of both instruments (H8 and H9), similar to the previous works (Gong et al., 2004; Teo, 2009); however, self-efficacy had almost no effect on perceived usefulness of technology in the model (H6 and H7). In addition to self-efficacy, we assumed that self-regulation was another external variable affecting students' perceptions. It was noted that students' initial self-regulation could determine the perceived usefulness (H11 and H12) while not directly influencing perceived ease of use (H10 and H13). The findings emerged that students' confidence in technology-enhanced environments tended to determine how they felt effortless in using technology. In contrast, students' self-regulation tended to affect how they consider technology could help them learn effectively. Self-efficacy is a student's belief in their ability to perform and learn effectively in a technology-enhanced environment. We found a strong relationship between self-efficacy and perceived ease of use, suggesting that students confident in their ability to learn well in a technology-enhanced environment may find it easier to use technology instruments but did not necessarily find them more useful. Nevertheless, interestingly, this belief did not extend to perceived usefulness, suggesting that while students may find it easy to use technological tools, they did not necessarily see them as improving their learning outcomes or effectiveness. Contrastingly, self-regulation was found to affect perceived usefulness but had less influence on perceived ease of use, implying that students with higher self-regulation, those capable of managing their learning processes effectively, tended to appreciate the usefulness of technology in supporting their learning. Students found the value of these tools in their learning process (e.g., planning, monitoring, and adjusting strategies), even though they did not necessarily find them easier to use.

The results of hypotheses H14, H15, H16, and H17 revealed that students' perceptions of web-based self-assessment standards played a pivotal role in determining self-regulation outcomes. In contrast, merely providing access to ICT for searching for information online did not significantly impact self-regulation outcomes. In this study, students utilized web-based self-assessment standards to evaluate their decision-making and argumentation quality on energy issues. Self-assessment, a metacognitive strategy, has been shown to influence academic performance (Brown & Harris, 2013) and students' SRL (Panadero et al., 2017). These findings drew attention to the importance of the specific affordances of technology as critical predictors, highlighting the need to move beyond basic technology integration and focus on incorporating tools specifically designed to support and enhance self-regulation.

We found three direct and four indirect paths determined self-regulation outcomes in the technology-enhanced SRL context. In other words, the direct relationship was crucial, where the increase in initial self-regulation, perceived usefulness and ease of use of web-based self-assessment standards corresponded to a subsequent rise in post-intervention self-regulation. Still, the direct paths were not the only determinants of outcomes. We also needed to account for indirect paths contributing to the overall effect on PostSR: SE → PreSR → PostSR, SE → PreSR → SASPU→ PostSR, SE → SASPEU → PostSR, and SE → SASPEU → SASPU → PostSR. We highlighted that although self-efficacy did not affect post-intervention self-regulation directly, affecting through the intermediary roles of initial self-regulation, perceived usefulness and ease of use of web-based self-assessment standards. These findings suggested the interconnected associations of self-efficacy, perceptions of technology, self-regulation outcomes, and the potential for self-efficacy to indirectly impact self-regulation by shaping perceptions of technology.

The substantive findings from this study represent only a portion of the contributions made by this manuscript to the existing literature. We also wish to emphasize our methodological approach. As van de Schoot et al. (2017) highlighted, Bayesian statistics have been employed across various subfields within psychology and related disciplines. Consequently, we utilized Bayesian estimation techniques to delve into the relationship between self-regulation and technology-enhanced environments, further enriching our understanding of this connection. B-SEM proved to be a vital approach in our study, as it allowed us to detect a marginal factor affecting PostSR, which might have otherwise been overlooked considering the p-value. Bayesian methods, in general, are powerful tools in statistical analysis, providing greater flexibility, enabling more nuanced interpretations, and allowing us to capture the uncertainty of assessments. In our study, applying B-SEM was instrumental in uncovering the somewhat marginal but still influential effect of SASPEU on PostSR. In this case, the posterior standard deviation, standing for the spread of the distribution, was quite large (0.11), suggesting considerable uncertainty around the estimate. This uncertainty was reflected in the 95% confidence interval, ranging from -0.03 to 0.40. The confidence interval included 0, suggesting that there was a possibility that there might be no effect of SASPEU on PostSR. However, the probability of direction value of 96% indicated that the effect of SASPEU on PostSR was positive in 96 out of 100 draws from the posterior distribution, suggesting that while there was some uncertainty, the bulk of the evidence points to a positive relationship between SASPEU and PostSR. This finding was not as evident when we employed ML-SEM because of the p-value, highlighting the importance of Bayesian estimation when dealing with complex models and nuanced relationships. The Bayesian approach provided a more comprehensive and robust view of the relationships in our structural model. By estimating the full posterior distributions of the parameters, we were able to assess the uncertainty surrounding our estimates more accurately, which is crucial in studies involving psychology, where variability is a common occurrence. We highly recommend the use of B-SEM in future studies, especially those dealing with complex models and relationships, where the additional insights gained through this approach can significantly enrich the overall findings.

We found it important to offer notes about the scope and limitations of this study. Firstly, the participants for our research were recruited from a city in Northern Taiwan, a region known for its strong emphasis on technological literacy. Consequently, these participants already possessed a fundamental proficiency in using technology for learning purposes. Therefore, these findings may not be generalizable to populations with different digital literacy levels or in different cultural contexts. Secondly, we adopted self-reported questionnaires to gauge factors of interest. Notably, self-reported data is susceptible to social desirability bias (Fisher, 1993), whereby participants may answer questions in a manner they believe to be socially acceptable or favorable rather than providing genuine, accurate responses. Therefore, future research could seek for other data to verify the findings of this study. It would be worthwhile to consider incorporating other methodologies in future studies, such as observations or interviews, to supplement self-reported data and provide a more comprehensive picture of the factors influencing self-regulation in technology-enhanced environments. Finally, while our model accounted for a significant portion of the variance in self-regulation outcomes, there may be other influential factors that were not considered in this study. Other individual attributes, such as motivation, personality traits, or learning styles, could also impact self-regulation and perceptions of technology. Future research should aim to identify and incorporate these additional variables to refine our understanding of technology-enhanced SRL further.

## 6. Conclusions

The present study sought to identify the factors influencing students' self-regulation outcomes in technology-enhanced learning environments. We discovered the direct and indirect effects on post-intervention self-regulation. Our findings thus provided a comprehensive understanding of technology-enhanced learning environments, illustrating the intricate interplay between self-efficacy, technology perceptions, and self-regulation. This research revealed the importance of nurturing self-efficacy and initial self-regulation skills while offering students favorable technology to optimize self-regulation outcomes. We found that self-efficacy might not influence post-intervention self-regulation directly while influencing indirectly via

4perceived the use of technology. Students' perceptions of SAS strongly influenced their post-intervention self-regulation. In contrast, merely providing students with ICT tools for online information retrieval did not necessarily enhance self-regulation. These findings implied that the mere provision of technology is insufficient to guarantee enhanced learning outcomes. Instead, the nature of the technology, specifically the capacity to support SRL, appears to be the critical determinant. Furthermore, this study highlighted Bayesian estimation in uncovering complex relationships and subtle effects, thereby demonstrating its potential to enrich research findings. The posterior distribution in Bayesian analysis provided valuable insights into the most credible values of parameters and their associated uncertainty. In contrast, ML-SEM tended to focus on p-values, which offer a binary view of significance but fail to capture the richness and uncertainty inherent in the data. Consequently, we advocate for broader adoption of this methodological approach in future studies, particularly those exploring intricate models and relationships. This study significantly advances our understanding of how technology-enhanced learning environments impact students' self-regulated learning. It underscores the need for developing pragmatic tools specifically tailored to reinforce self-regulation in learning.

7. **Acknowledgements**

This work was financially supported by the National Science Council of Taiwan under contracts the MOST 111-2410-H-003-032-MY3, the NSTC 111-2423-H-003-004, MOST 110-2511-H-003 -027 -MY2 and the "Institute for Research Excellence in Learning Sciences" of National Taiwan Normal University (NTNU) from The Featured Areas Research Center Program within the framework of the Higher Education Sprout Project by the Ministry of Education (MOE) in Taiwan.

36

**Appendix A**

Table1A

*Survey instrument*

| Item | Description |
| --- | --- |
| SR1 | Before learning new content in a course, I usually browse through the learning material to understand how it is organized. |
| SR2 | During the course, I set goals for myself and plan what I should do at each learning stage. |
| SR3 | In the course, I check to confirm my learning performance. |
| SR4 | In the course, I ask myself questions to confirm my level of understanding. |
| SR5 | In the course, when encountering confusing learning content, I make every effort to clarify and understand it. |
| SE1 | I am confident that I can use technology to assist in self-regulated learning and master the most complex learning content provided by the teacher. |
| SE2 | I am confident that I can use technology to assist in self-regulated learning and perform well in assignments and exams in biology courses. |
| SE3 | Considering the difficulty of the course, the teacher, and my skills, I can use technology to assist in self-regulated learning and perform well in this course. |
| ICTPU1 | Using ICT tools (such as tablets, Google, etc.) can improve my self-regulated learning in SSI. |
| ICTPU2 | Using ICT tools (such as tablets, Google services, etc.) can increase the efficiency of my self-regulated learning in SSI. |
| ICTPU3 | ICT tools (such as tablets, Google, etc.) allow me to complete self-regulated learning tasks more effectively. |
| ICTPEU1 | I find the various features of ICT tools (such as tablets, Google, etc.) clear and easy to understand and use. |
| ICTPEU2 | Operating ICT tools (such as tablets, Google, etc.) makes me feel relaxed and not burdened. |

| | | |
|---|---|---|
| ICTPEU3 | I can easily remember how to use ICT tools (such as tablets, Google, etc.) for self-regulated learning tasks. | |
| SASPU1 | The web-based self-assessment and standards provided in class can improve my monitoring ability. | |
| SASPU2 | The web-based self-assessment and standards provided in class helps me more accurately judge my understanding of argumentation. | |
| SASPU3 | The web-based self-assessment and standards provided in class can improve my argumentation performance. | |
| SASPEU1 | The web-based self-assessment and standards provided in class are clear and easy to understand. | |
| SASPEU2 | The web-based self-assessment and standards provided in class are easy to use. | |
| SASPEU3 | I can easily use the web-based self-assessment and standards to judge my understanding or performance without effort. | |

Table 2A

*The results of the measurement model based on ML-estimation*

| Construct | Item | Factor loading(>0.7)[a] | CR (>0.7)[a] | AVE (>0.5)[a] |
|---|---|---|---|---|
| PreSR | PreSR1 | 0.736 | 0.877 | 0.590 |
| | PreSR2 | 0.776 | | |
| | PreSR3 | 0.840 | | |
| | PreSR4 | 0.796 | | |
| | PreSR5 | 0.682 | | |
| PostSR | PostSR1 | 0.816 | 0.901 | 0.644 |
| | PostSR2 | 0.784 | | |
| | PostSR3 | 0.873 | | |
| | PostSR4 | 0.754 | | |
| | PostSR5 | 0.794 | | |
| SASPU | SASPU1 | 0.888 | 0.924 | 0.801 |
| | SASPU2 | 0.920 | | |
| | SASPU3 | 0.880 | | |
| SASPEU | SASPEU1 | 0.867 | 0.885 | 0.719 |
| | SASPEU2 | 0.838 | | |
| | SASPEU3 | 0.839 | | |
| ICTPU | ICTPU1 | 0.847 | 0.925 | 0.805 |
| | ICTPU2 | 0.920 | | |
| | ICTPU3 | 0.919 | | |
| ICTPEU | ICTPEU1 | 0.862 | 0.893 | 0.735 |
| | ICTPEU2 | 0.875 | | |
| | ICTPEU3 | 0.837 | | |
| SE | SE1 | 0.807 | 0.900 | 0.749 |
| | SE2 | 0.898 | | |
| | SE3 | 0.896 | | |

Note: [a] Indicates an acceptable level of reliability or validity in line with Fornell and Larcker (1981) and Kline (2011). SE stands for self-efficacy.